\documentclass[letterpaper,fleqn,usenatbib]{mnras}

\usepackage{newtxtext,newtxmath}
\usepackage[T1]{fontenc}
\usepackage{ae,aecompl}

\usepackage{graphicx}	
\usepackage{amsmath}	
\usepackage{amssymb}	

\title[RZ~Psc is a binary]{A low-mass stellar companion to the young variable star RZ~Psc}

\author[G. M. Kennedy et al.]{
Grant M. Kennedy,$^{1,2}$\thanks{E-mail: g.kennedy@warwick.ac.uk}
Christian Ginski,$^{3,4}$
Matthew A. Kenworthy,$^{4}$
Myriam Benisty,$^{5,6}$
\newauthor
Thomas Henning,$^{7}$
Rob G. van Holstein,$^{4,8}$
Quentin Kral,$^{9}$
Fran\c cois M\'{e}nard,$^{5}$
Julien Milli,$^{5}$
\newauthor
Luis Henry Quiroga-Nu\~nez,$^{10,11}$\thanks{Jansky Fellow of the National Radio Astronomy Observatory}
Christian Rab,$^{12}$
Tomas Stolker,$^{13}$
Ardjan Sturm$^{4}$
\\
$^{1}$Department of Physics, University of Warwick, Gibbet Hill Road, Coventry, CV4 7AL, UK\\
$^{2}$Centre for Exoplanets and Habitability, University of Warwick, Gibbet Hill Road, Coventry CV4 7AL, UK\\
$^{3}$Sterrenkundig Instituut Anton Pannekoek, Science Park 904, 1098 XH Amsterdam, The Netherlands\\
$^{4}$Leiden Observatory, Leiden University, PO Box 9513, 2300 RA Leiden, The Netherlands\\
$^{5}$Univ. Grenoble Alpes, CNRS, IPAG, F-38000 Grenoble, France\\
$^{6}$Unidad Mixta Internacional Franco-Chilena de Astronom\'ia (CNRS, UMI 3386), Departamento de Astronom\'ia, Universidad de Chile,\\ Camino El Observatorio 1515, Las Condes, Santiago, Chile\\
$^{7}$Max Planck Institute for Astronomy, K\"onigstuhl 17, 69117 Heidelberg, Germany\\
$^{8}$European Southern Observatory (ESO), Alonso de C\'ordova 3107, Vitacura, Casilla 19001, Santiago, Chile\\
$^{9}$LESIA, Observatoire de Paris, Universit\'e PSL, CNRS, Sorbonne Universit\'e, Univ. Paris Diderot, Sorbonne Paris Cite\'e, 5 place Jules Janssen,\\ F-92195 Meudon, France\\
$^{10}$National Radio Astronomy Observatory, Array Operations Center, Socorro NM 87801 USA\\
$^{11}$Department of Physics and Astronomy, University of New Mexico, Albuquerque NM 87131 USA\\
$^{12}$Kapteyn Astronomical Institute, University of Groningen, PO Box 800, 9700 AV Groningen, The Netherlands\\
$^{13}$Institute for Particle Physics and Astrophysics, ETH Zurich, Wolfgang-Pauli-Strasse 27, 8093 Zurich, Switzerland\\
}

\date{Accepted 2020 May 12. Received 2020 April 17; in original form 2020 February 18}

\pubyear{2020}

\begin{document}
\label{firstpage}
\pagerange{\pageref{firstpage}--\pageref{lastpage}}
\maketitle

\begin{abstract}
RZ~Psc is a young Sun-like star with a bright and warm infrared excess that is occasionally dimmed significantly by circumstellar dust structures. Optical depth arguments suggest that the dimming events do not probe a typical sight line through the circumstellar dust, and are instead caused by structures that appear above an optically thick mid-plane. This system may therefore be similar to systems where an outer disk is shadowed by material closer to the star. Here we report the discovery that RZ~Psc hosts a $0.12\,M_\odot$ companion at a projected separation of 23\,au. We conclude that the disk must orbit the primary star. While we do not detect orbital motion, comparison of the angle of linear polarization of the primary with the companion's on-sky position angle provides circumstantial evidence that the companion and disc may not share the same orbital plane. Whether the companion severely disrupts the disc, truncates it, or has little effect at all, will require further observations of both the companion and disc.
\end{abstract}




\section{Introduction}\label{s:intro}

RZ~Psc is a young star that shows irregular photometric dimming events that are attributed to circumstellar dust transiting across the face of the star. Such systems promise a window on the clumpiness of the inner regions of circumstellar discs near the epoch of planet formation \citep[e.g.][]{2013A&A...553L...1D}, and combined with ancillary information (e.g. dust thermal emission, outer disc imaging) may yield broader insights into disc vertical structure and geometry \citep[e.g.][]{2017RSOS....460652K,2016MNRAS.462L.101A,2020MNRAS.492..572A}. Similarly, clumpy and/or misaligned inner discs can cast shadows on outer disc regions \citep[e.g.][]{2012ApJ...754L..31C,2015ApJ...798L..44M,2018ApJ...868...85P}. As noted by \cite{2017ApJ...849..143S}, shadowed-disc systems may be very similar to those where the star is seen to be dimmed from Earth, and simply observed with a more face-on geometry. Given that perturbations of inner disc regions by planetary or stellar companions has been proposed for shadowed-disc systems \citep[e.g.][]{2018ApJ...868...85P}, the same may be true for variable systems such as RZ~Psc.

RZ~Psc has been classed as an "UXor" \citep[e.g.][]{1999AstL...25..243R}, which are normally young Herbig Ae/Be stars that show sporadic photometric dimming events that last for days to weeks \citep{1994AJ....108.1906H}. UXors also tend to show an increased degree of linear polarization and reddening during typical dimming events \citep[e.g.][]{1994ASPC...62...63G,1988SvAL...14...27G}. The circumstellar dust interpretation posits that the linear polarization increases because the fraction of the total emission that is contributed by dust-scattered light increases when the star is dimmed, and that the reddening is caused by wavelength-selective extinction. 
While other scenarios may explain some UXors \citep{1999AJ....118.1043H}, RZ~Psc is consistent with the circumstellar dust scenario, for several reasons.

Primarily, RZ~Psc shows both the reddening behaviour and an increased degree of linear polarization during dimmings \citep{2003ARep...47..580S}. Using the timescale of RZ~Psc's dimming events, from which a transverse velocity can be estimated, this dust is estimated to lie at roughly 1\,au \citep[5\,mas at the distance of 196\,pc,][]{2010A&A...524A...8G,2018A&A...616A...1G}. This system is also seen to host a bright mid-infrared (IR) excess, inferred to be thermal emission from a warm circumstellar dust population \citep{2013A&A...553L...1D}. There is no far-IR excess, which rules out a bright outer disc, but the outer radius could nevertheless be several tens of au \citep{2017RSOS....460652K}. While neither of these properties are unusual for an UXor, what marks RZ~Psc as unusual is that its probable age is a few tens of Myr, older than other UXors, and Herbig Ae/Be stars in general \citep[which have ages $\lesssim$10\,Myr,][]{2013AstL...39..776P,2010A&A...524A...8G}. Also, the spectral type of K0V is later than typical UXors, and more similar to the so-called "dippers" \citep[which show more frequent dust-related dimming behaviour, typically on $\sim$day timescales, see e.g.][]{2011ApJ...733...50M,2014AJ....147...82C,2016ApJ...816...69A,2018MNRAS.476.2968H}. In addition to the short-term dimming events, RZ~Psc also shows a long-term sinusoidal photometric variation with a 12\,year period and 0.5\,mag amplitude.

Synthesising these observations, \citet{2013A&A...553L...1D} suggested that the dust populations causing the dimming and IR excess are one and the same, and given the probable age that this dust resides in a young and massive analogue of the Solar system's asteroid belt.
The long-term variation was considered largely independent of the inner dust population, perhaps caused by a varying line of sight through an outer disc that is warped by a companion.

This picture was reconsidered by \citet{2017RSOS....460652K}, based primarily on a simple optical depth argument. The large fractional dust luminosity ($L_{\rm disc}/L_\star \approx 0.07$) implies that 7\,\% of the starlight is captured and re-emitted thermally by the dust. That is, as seen from the star about 7\,\% of the sky is covered by dust.
RZ~Psc is not seen to be significantly reddened \citep{2000ARep...44..611K}, implying that the dust must either have a large scale height (i.e. has a more shell-like structure), or that our line of sight to the star does not probe a typical sight line through the dust cloud. The increased degree of linear polarization during deep dimming events however disfavours a near-spherical distribution of dust around the star. \citet{2017RSOS....460652K} therefore concluded that the dimming events are the result of occultations by dust structures that rise above an optically thick disc mid-plane, highlighting the alternative possibility that RZ~Psc hosts a relatively old and evolved protoplanetary disc \citep[a related scenario is that material is falling towards the star, so appears above the mid-plane in projection, e.g.][]{1996A&AS..120..157G}.
Recent work finds that RZ~Psc shows a weak H$\alpha$ accretion signature, which is more consistent with a long-lived protoplanetary disc scenario \citep{2017A&A...599A..60P,2019A&A...630A..64P}.

Here we present the discovery that RZ~Psc is a binary with a sky-projected separation of 23\,au, based on high-contrast imaging observations with VLT/SPHERE (\citealt{2019A&A...631A.155B}). While RZ~Psc shows no evidence for an outer disc, this companion might be truncating and/or warping parts of the disc in a way that makes the inner regions of this system similar to shadowed-disc systems. We present our observations in Section \ref{s:obs}, and discuss the possible implications in Section \ref{s:disc}.

\section{Observations and Results}\label{s:obs}

\begin{table}
 \centering
 \caption{Observing setup and average observing conditions for all SPHERE/IRDIS observation epochs. DIT is the detector integration time for each frame, and $\langle\tau_0\rangle$ is the average coherence time of the atmosphere. Only the 16-08-2019 observation did not use a coronagraph.}
  \begin{tabular}{@{}cccccc@{}}
  \hline 
 Epoch & Filter & DIT [s] & \# of frames & Seeing [\arcsec{}] & $\langle\tau_0\rangle$ [ms] \\
 \hline
01-10-2018	& BB\_H & 64 & 32 & 0.81 & 3.9 \\
16-08-2019 & BB\_H & 0.84 & 1536 & 0.63 & 4.7 \\
17-08-2019 & BB\_K$_s$ & 64 & 32 & 0.68 & 2.4 \\
19-08-2019 & BB\_K$_s$ & 64 & 32 & 0.58 & 3.5 \\
21-08-2019 & BB\_K$_s$ & 64 & 32 & 0.40 & 14.7 \\
23-08-2019 & BB\_K$_s$ & 64 & 32 & 0.75 & 4.3 \\

\hline\end{tabular}
\label{tab: observing_setup}
\end{table}

RZ~Psc was observed with SPHERE/IRDIS (\citealt{2008SPIE.7014E..3LD}) in its dual-band polarimetric imaging (DPI) mode (\citealt{2014SPIE.9147E..1RL,2020A&A...633A..63D,2020A&A...633A..64V}) at a total of six observation epochs. We show the instrument setup and integration time in Table~\ref{tab: observing_setup}. The first observation epoch in 2018 was performed in the field stabilized mode of the instrument, while all subsequent observations were performed in pupil stabilized mode to ensure high polarimetric throughput of the instrument (\citealt{2017SPIE10400E..15V}). We note that the field rotation due to the parallactic angle change in pupil stabilized mode was 10$^\circ$ or less in all observation epochs. All but the second observation epoch use the IRDIS apodized Lyot coronagraph \emph{YJH\_S} with a radius of 92.5\,mas (\citealt{2011ExA....30...39C}).

All data reduction was performed with the IRDAP pipeline (IRDIS Data reduction for Accurate Polarimetry, \citealt{2020A&A...633A..64V}). The final data products used for the analysis included the stacked total intensity images as well as the Stokes $Q$ and $U$ images. Due to the very small change in parallactic angle we did not perform angular differential imaging in total intensity on the data sets. 

\begin{figure*}
\center
\includegraphics[width=0.98\textwidth]{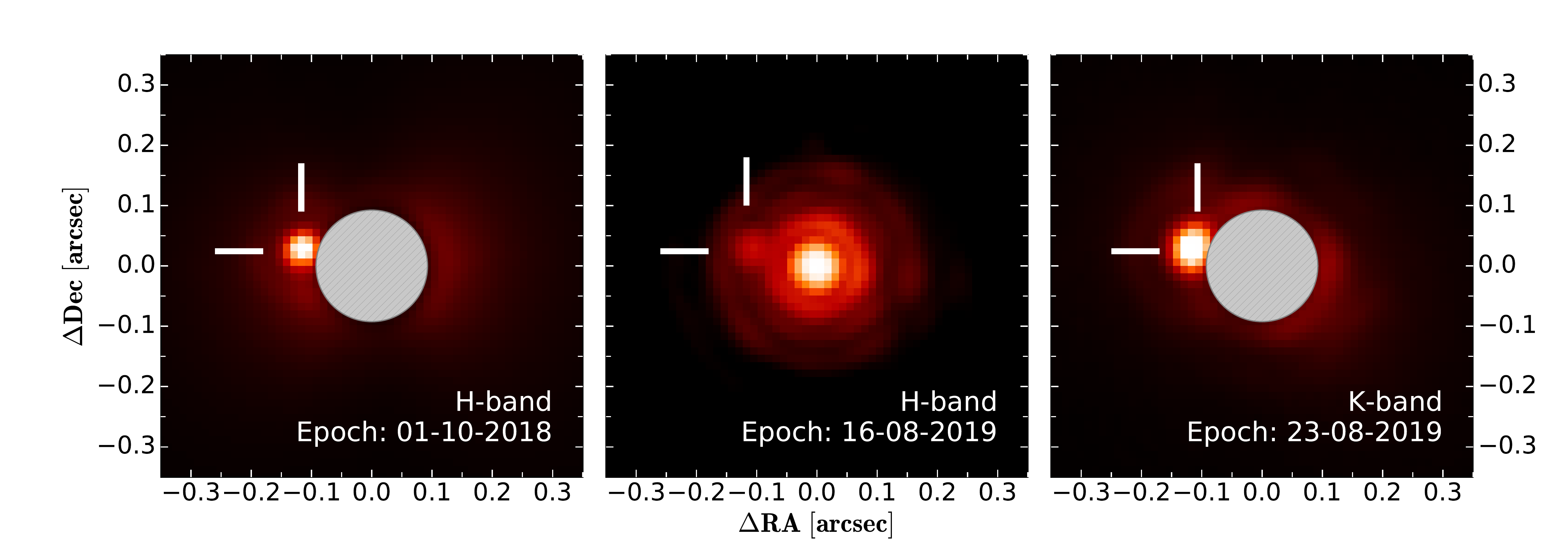} 
\caption{SPHERE/IRDIS observations of RZ~Psc~A's companion. Left and right images were taken with a coronagraph in place (marked by the grey hashed area). The data shown in the middle panel was taken without a coronagraph. The position of the new stellar companion is marked by the white bars. The filter of the observation is indicated in each panel as well as the observation date. The J2000 primary position is 01h09m42.05s +27d57m01.9s at epoch 2000.0.
} 
\label{fig:sphere_images}
\end{figure*}

For our analysis we used the observations taken in the H-band on 01-10-2018 and 16-08-2019, with and without the coronagraph respectively, as well as the observation taken in K$_s$-band on 23-08-2019. The remaining three K$_s$-band epochs suffer from non-ideal observing conditions as described in Appendix \ref{K_appendix}. The total intensity images reveal the presence of a close and faint companion in the system as shown in Figure~\ref{fig:sphere_images}.

\begin{figure}
\center
\includegraphics[width=0.48\textwidth]{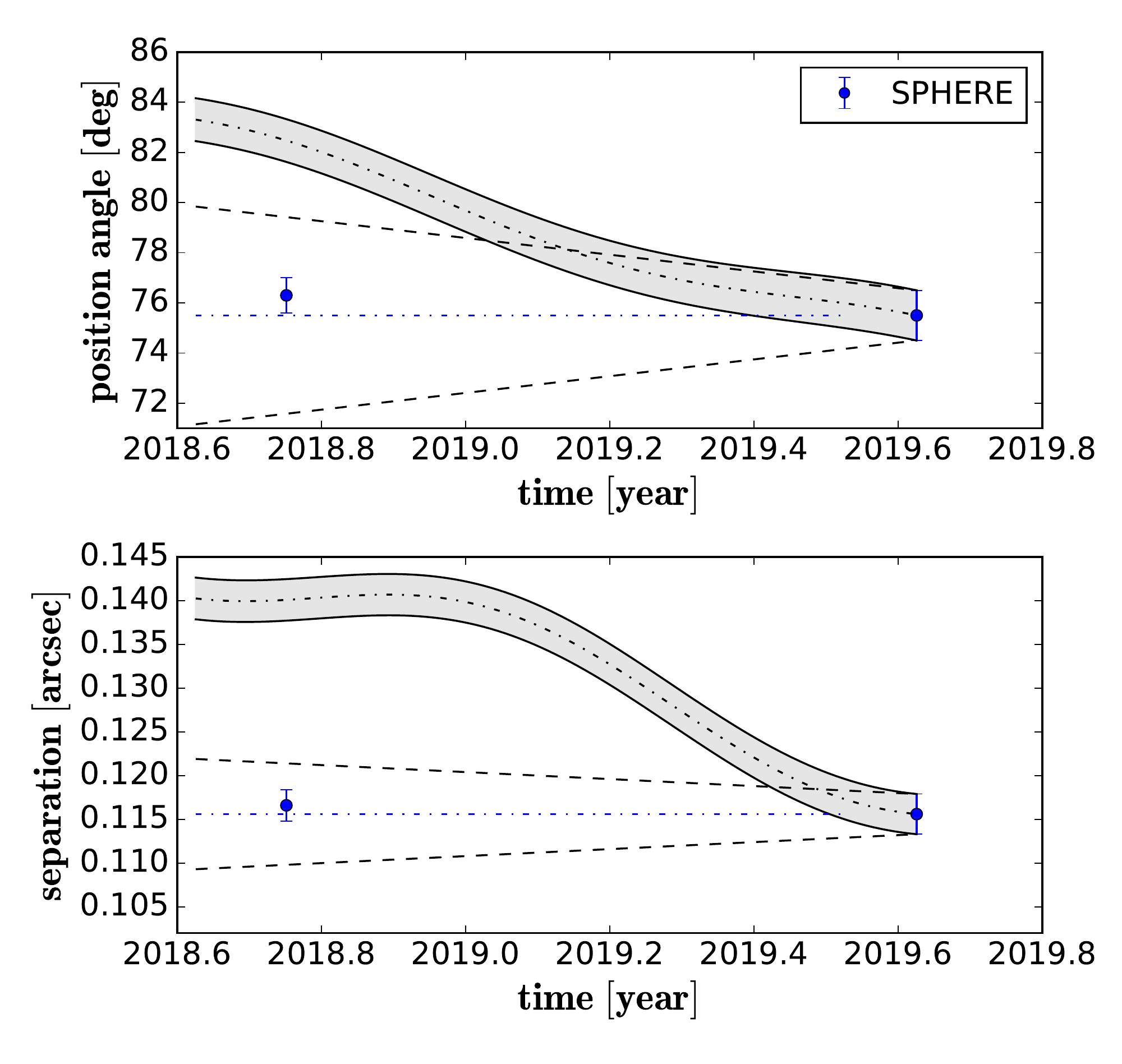} 
\vspace{-0.5cm}
\caption{Position angle and separation of the new companion relative to RZ\,Psc\,A over time. The grey ribbon area indicates the expected behavior for an unmoving background object, given RZ~Psc's proper motion of $\mu_\alpha =27.4 \pm 0.1$\,mas\,yr$^{-1}$, $\mu_\delta = 	-12.6 \pm 0.2$\,mas\,yr$^{-1}$ \citep{2018A&A...616A...1G}. The dashed lines show the maximal angle/distance a co-moving companion on a circular orbit could move, assuming face-on (upper panel), and edge-on (lower panel) orientations, and the stellar mass and distance given in section \ref{s:disc}.
} 
\label{fig:pm_plot}
\end{figure}

To determine the nature of the companion we extracted astrometry and photometry from all observation epochs. Since the companion position is strongly contaminated by the point-spread function (PSF) of RZ~Psc we first subtracted a 180$^\circ$ rotated version of the images from the images to approximate the stellar point spread function. For astrometric extraction we then fitted a Gaussian to the companion signal. Astrometric calibration of the images was performed with the standard values given in \cite{2016SPIE.9908E..34M}. The final astrometric results are given in Table~\ref{tab: astrometry}. Note that for the K$_s$-band epoch the companion signal is likely vignetted somewhat by the coronagraphic mask, which may influence the extracted position. This should be less problematic in the higher spatial resolution H-band image of the first observation epoch and the image taken without the coronagraph in the second observation epoch.

To extract the photometry of the companion we used apertures. The aperture radii were 3 detector pixels in the H-band images and 4 detector pixels in the K$_s$-band images,  approximating one resolution element in either band. In addition to the initial background subtraction, we used apertures with the same separation from the primary star at larger and smaller position angles than the companion position to determine the mean local residual background, which we subtract from the companion flux. All photometric measurements were performed relative to the primary star since no other calibrators were available. For the non-coronagraphic H-band epoch, both the star and the companion flux could be measured in the same image. To measure the stellar flux in the coronagraphic epochs we used flux calibration frames in which the primary star was shifted away from the coronagraph. These were taken with short integration times and, in the case of the H-band, a neutral density filter to prevent saturation. We used IRDIS \emph{ND\_1.0} neutral density filter, with a throughput ratio in H-band of 7.94. Since the companion is located close to the inner working angle of the coronagraph we have to take coronagraph throughput into account. For the H-band detailed measurements exist and are presented in a forthcoming publication by Wilby et al, in prep. Given the companion separation we used a throughput correction factor of 0.865. For the K$_s$-band no measurements exist for the \emph{YJH\_S} coronagraph. However, since the PSF full width at half maximum scales linearly with wavelength we extrapolated the throughput correction from shorter wavelengths and find a correction factor of 0.765. The extracted magnitude relative to the primary for each observation is given in Table \ref{tab: astrometry}

\begin{table}
 \centering
 \caption{Astrometry and photometry of the RZ\,Psc system, including 1$\sigma$ uncertainties, as extracted from our SPHERE/IRDIS observations. The apparent magnitude assumes that the primary has a magnitude of 9.84 in H-band, and 9.70 in K$_s$-band \citep[2MASS,][]{2006AJ....131.1163S}.}
  \begin{tabular}{@{}cccccc@{}}
  \hline 
 Epoch & Filter & Sep [mas] & PA [deg] & $\Delta$mag & app. mag \\
 \hline
 
01-10-2018	& H & 116.6$\pm$1.8 & 76.3$\pm$0.7 & 3.87$\pm$0.14 & 13.71$\pm$0.14\\
 16-08-2019	& H & 115.6$\pm$2.3 & 75.5$\pm$1.0 & 4.02$\pm$0.26 & 13.85$\pm$0.26\\
 23-08-2019 & K$_s$ & 120.8$\pm$1.8 & 74.9$\pm$0.7 & 3.81$\pm$0.20& 13.51$\pm$0.20\\

\hline\end{tabular}
\label{tab: astrometry}
\end{table}


All-Sky Automated Survey for Supernovae \citep{2017PASP..129j4502K} photometry on the days before and after our first epoch observation are consistent with being in the undimmed state. While photometry is not yet available for the more recent epochs, the consistency of the photometry in Table \ref{tab: astrometry} suggests that these observations were also during undimmed periods. Similarly, SED fitting that includes the 2MASS photometry used to derive the apparent magnitudes of the companion finds a well-fitting solution, indicating that the star was not significantly dimmed at the time of those observations \citep{2017RSOS....460652K}.

We do not detect a linear polarization signal from extended circumstellar dust in any of the observation epochs and can thus rule out the presence of significant amounts of small ($\mu$m-sized) dust particles down to the coronagraphic inner working angle of 92\,mas (18\,au). We do measure the linear polarization of the primary star in H-band (the K$_s$-band data are of not sufficient quality to provide a useful measurement). The first and second epochs yield $0.5 \pm 0.1\,\%$ at PA $52 \pm 7^\circ$ and $0.56 \pm 0.09\,\%$ at PA $55 \pm 3^\circ$ respectively. 
Due to the contamination by the residual stellar halo in Stokes Q and U images a measurement of the companion polarization is difficult. Using the same aperture photometry technique as described previously, we estimate that the companion's degree of linear polarization is less than 0.2\,\% in the H-band.

\section{Discussion and Conclusions}\label{s:disc}

Several tests can be made to verify association of the companion with RZ~Psc. First, following the approach of \citet{2014A&A...566A.103L} we used the \texttt{TRILEGAL} model of the Milky Way \citep{2005A&A...436..895G} to estimate the probability of a background star as bright or brighter than the detected companion appearing within 120\,mas of RZ~Psc, which yields $1.7 \times 10^{-6}$. Second, Figure \ref{fig:pm_plot} shows that the companion is co-moving with RZ~Psc on the sky and that a stationary background source is ruled out (5.6$\sigma$ in PA, 7.9$\sigma$ in separation). Third, while a background source is not necessarily stationary, the distance required for a K0-type giant (e.g. Pollux) to have the same magnitude as the companion is about 9\,kpc, and at this distance fewer than 1 in 10,000 stars in Gaia Data Release 2 \citep{2016A&A...595A...1G,2018A&A...616A...1G} have a proper motion as high as RZ~Psc. Also, given the few-degree uncertainty on the companion PA, the probability that a background source with the same proper motion amplitude would also be moving in the same direction as RZ~Psc is $\sim$1\,\%. We therefore conclude that the companion is almost certainly associated with RZ~Psc.

Using the distance to RZ~Psc of 196\,pc, the absolute H-band magnitude of the companion is $7.3 \pm 0.1$ and the H-K$_s$ colour is $0.2 \pm 0.3$, which are consistent with a mid-M field star spectral type \citep[e.g.][]{2020MNRAS.492..431B}. If we assume an age of 20\,Myr for RZ~Psc \citep{2019A&A...630A..64P}, then using the BT-Settl isochrones \citep{2012RSPTA.370.2765A} we find a companion mass of 0.12$\pm$0.01\,$M_\odot$ for the H-band and 0.11$\pm$0.02\,$M_\odot$ for the K$_s$-band. The systematic uncertainty in the mass due to an uncertain age is of course larger (on the order of 0.05\,$M_\odot$ if the age changes by 10\,Myr). 

Assuming a circular and face-on orbit, the expected azimuthal orbital motion of the companion at 23\,au (assuming 196\,pc and an approximate primary mass for a K0V-type star of 0.9\,$M_\odot$) is about 6\,mas\,yr$^{-1}$. Our observations are on the verge of providing constraints on the orbit, and similarly precise observations beyond 2020 will constrain the orbit regardless of whether relative motion is seen. For now however, we cannot derive meaningful constraints because i) the object's motion is not sufficiently high to rule out low eccentricity or high inclination orbits \citep{2015MNRAS.448.3679P}, and ii) the lack of motion between epochs remains consistent with an unbound object; the 3$\sigma$ upper limit on the sky-projected velocity is 1.6\,au\,yr$^{-1}$, only slightly less than the escape velocity at the projected separation of 23\,au of 1.8\,au\,yr$^{-1}$ (and the actual separation is likely larger, so the escape velocity lower). While a close encounter between unbound objects therefore remains a possibility, this is also unlikely because the interaction at the observed separation would last of order hundreds of years, which is a tiny fraction of the stellar age.


Our H-band linear polarization measurements are similar, in both magnitude and angle, to those in the optical when the star is not significantly dimmed and after interstellar polarization has been subtracted, suggesting that most of the polarization in the near-IR comes from light scattered off the circumstellar disc \citep[optical measurements in $V$ and $R$ yield 0.68\,\% at 48$^\circ$ and 0.73\,\% at 49$^\circ$ respectively,][]{2003ARep...47..580S}.
For most disk orientations, the unresolved linear polarisation angle is perpendicular to the disk position angle, though can become parallel in highly optically thick and near to edge-on cases \citep{1992ApJ...395..529W,1993ApJ...402..605W}. Given that RZ~Psc is not significantly reddened when undimmed, the latter seems unlikely. The angle of linear polarization of 53$^\circ$ is therefore more similar to the companion's on-sky PA of 76$^\circ$ than expected, and may mean that the disk plane and companion orbital plane are misaligned. The companion's orbit remains unconstrained however, so this discrepancy is largely suggestive, and would be aided by observations that establish the disk orientation for comparison with the angle of polarisation. 


What does this discovery mean for the interpretation of RZ~Psc, and the nature of the UX~Ori-like dimming events? Firstly, the companion is sufficiently faint (i.e. $\approx$40$\times$ fainter than RZ~Psc~A in H-band) that if it were the object being dimmed, the effect on the light curve at any wavelength would be negligible. Similarly, the companion's luminosity ($\ll$2.5\,\% of RZ~Psc~A's luminosity) is less than the IR excess (7\,\%), so cannot solely heat the observed circumstellar dust. Therefore, the circumstellar dust orbits the primary.


It seems inevitable that the occulting structures causing the dimming of RZ~Psc~A reside above a disc mid-plane that has significant radial optical depth. Our data strengthen this view, the reason being that the PA of linear polarization measured here is very similar to that measured by \citet{2003ARep...47..580S}. If the dust were in a roughly spherical cloud and the linear polarization arises due to inhomogeneities in this structure, then the PA would not be expected to be similar for measurements made decades apart. Nevertheless, it is possible that the companion strongly influences the disc dynamics, for example via truncation at $\sim$1/2 to 1/3 of the companion's semi-major axis (truncation seems more likely than gap formation, given the lack of far-IR excess). If the disc still retains a significant gas mass this influence may also manifest as a warp or spiral structures that pull material out of the disc plane and into our line of sight to the star. If the disc is gas-poor, the companion may have recently destabilised a newly formed planetary system or planetesimal belt, which is now colliding and producing a significant mass of dust, some of which passes between us and the star. In either case a companion could cause some portions of the disc to precess slowly, or otherwise cause the 0.5\,mag variation seen with a 12\,year period \citep[e.g. as discussed by][]{2013A&A...553L...1D}.

The specifics of such interactions are unclear; while we find circumstantial evidence that the disc and companion have different orbital planes, this would be best quantified by constraining the companion's orbit (e.g. with VLT/SPHERE), and attempting to derive constraints on the dust geometry (e.g. with VLTI/MATISSE). ALMA observations may provide useful information on material beyond a few au, e.g. a continuum or CO detection could reveal the presence of an outer disc, or a non-detection suggest that the disc extent is largely limited to that already seen in the mid-IR. Such characterisation may also shed light on the long-term variation.

What are the wider implications of this discovery, if any? One is that the inner companion theory for shadowed outer disc systems seems credible, if RZ~Psc is indeed a more highly inclined version of such systems. Another is that while young systems with extreme warm IR excesses are commonly interpreted as showing evidence for terrestrial planet formation \citep[e.g.][]{2004ApJ...602L.133K,2010ApJ...717L..57M,2012ApJ...751L..17M}, the possible influence of more massive companions should be considered.
While further work is clearly needed, RZ~Psc provides evidence that non-planetary mass companions may play an important role in the evolution of young planetary systems.

\section*{Acknowledgements}

We thank the referee for a constructive report. GMK is supported by the Royal Society as a Royal Society University Research Fellow. CG and CR acknowledge funding from the Netherlands Organisation for Scientific Research (NWO) TOP-1 grant as part of the research programme “Herbig Ae/Be stars, Rosetta stones for understanding the formation of planetary systems”, project number 614.001.552. FM acknowledges funding from ANR of France under contract number ANR-16-CE31-0013.








\appendix

\section{Additional K$_s$-band epochs}
\label{K_appendix}

\begin{figure}
\center
\includegraphics[width=0.48\textwidth]{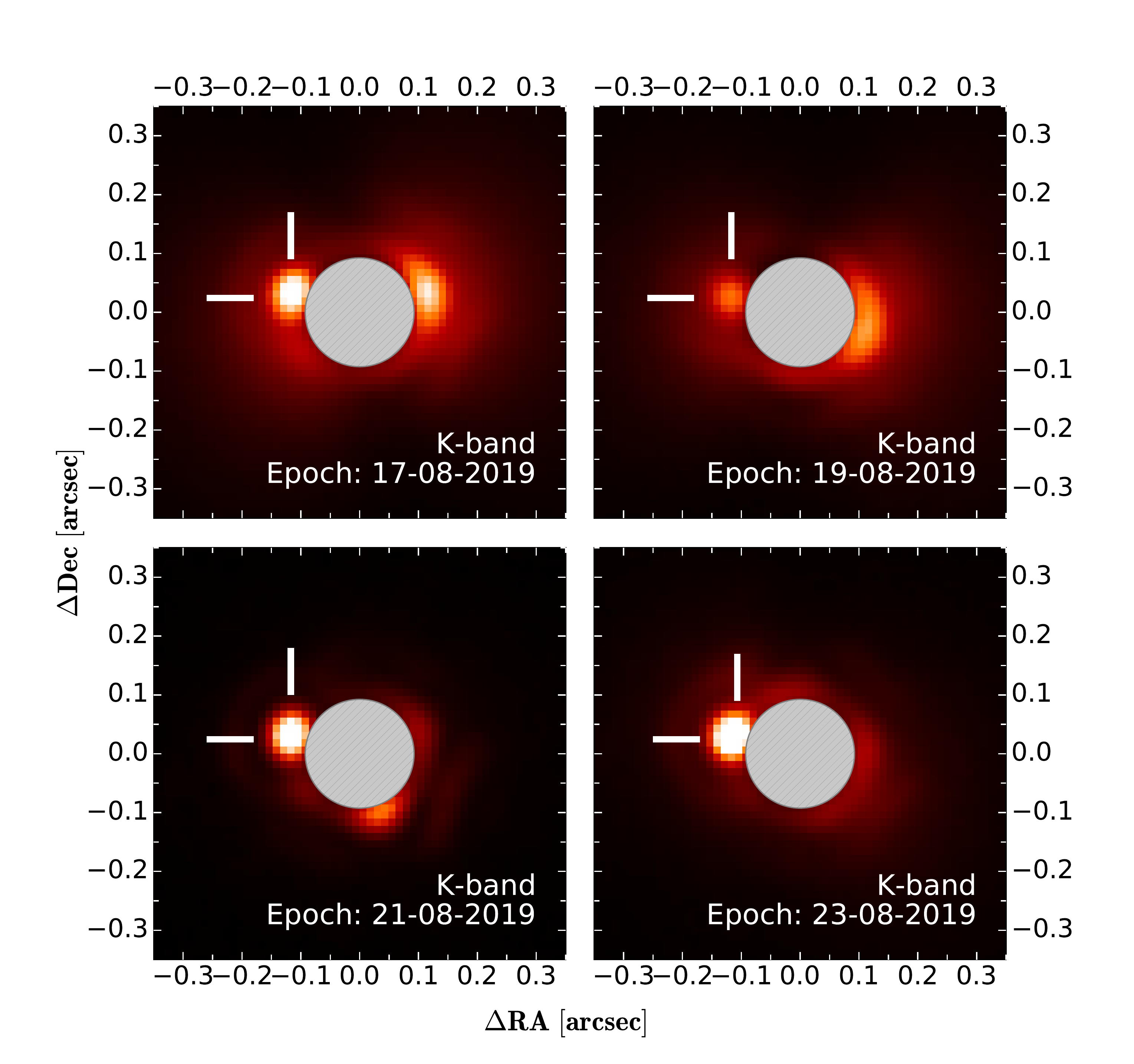} 
\vspace{-0.4cm}
\caption{Images of all four K$_s$-band observations (see Table \ref{tab: observing_setup}). Only epoch 4 (23-08-2019) is used in our analysis. All data sets are shown with the same stretch of the color bar.} 
\label{fig:k}
\end{figure}

Figure \ref{fig:k} shows the four K$_s$-band images, of which only epoch 4 was used in the final analysis. Epoch 1 suffers from a strong wind-driven halo \citep{2018A&A...620L..10C}, which results in asymmetric bleeding of the stellar halo from behind the coronagraph, making estimation of the background at the location of the companion uncertain. Epoch 2 suffers similar issues, but is markedly worse. Epochs 3 and 4 are better, but epoch 3 still suffers from an asymmetric PSF, this time due to the low wind effect \citep{2016SPIE.9909E..16S,2018SPIE10703E..2AM}. Epoch 4 does not suffer either of these issues.

\bsp	
\label{lastpage}
\end{document}